\documentclass[twocolumn,showpacs,preprintnumbers,amsmath,amssymb]{revtex4}
\usepackage{graphicx}
\usepackage{dcolumn}
\usepackage{bm}
\usepackage{amsmath,amssymb,amsfonts}
\begin{document}

\title{Prolate horizons and the Penrose inequality}

\author{Benjamin K. Tippett}
 \altaffiliation[Also at ]{Department of Mathematics and Statistics, UNB}
 \email{v5pv3@unb.ca}

\date{\today}
\begin{abstract}
 The Penrose inequality has so far been proven in cases of spherical symmetry and in cases of zero extrinsic curvature. The next simplest case worth exploring would be non-spherical, non-rotating black holes with non-zero extrinsic curvature.  Following Karkowski \emph{et al.}'s construction of prolate black holes, we define initial data on an asymptotically flat  spacelike 3-surface with nonzero extrinsic curvature that may be chosen freely. This gives us the freedom to define the location of the apparent horizon such that the Penrose inequality is violated. We show that the dominant energy condition is violated at the poles for all cases considered. 

\end{abstract}

\maketitle

\section{\label{sec:level1}Introduction}

The Penrose inequality is a conjectured upper bound relating the area of a black hole to the total mass of the spacetime. Should it ever be generally proven, we would be able to make stronger assumptions when studying the general properties of classical black holes and modelling gravitational collapse. Conversely, a carefully constructed counterexample could serve to invalidate the Cosmic Censorship Conjecture \cite{penrose1},\cite{wald1}. Either way, a deeper understanding of the Penrose inequality will prove fruitful.

In studying the evolution of a black hole, the object of interest is frequently the \emph{trapping horizon}. This is a 3-surface consisting of the outermost, continuously evolving \emph{trapped surface}, which tells us where an observer travelling at the speed of light will neither fall into the black hole, nor escape from it. As matter falls through and becomes trapped inside the black hole, the trapping horizon will expand until the black hole runs out of food and the \emph{trapped surface} becomes an \emph{event horizon}. The area of a trapped surface at any time is therefore smaller than the area of the event horizon \cite{hayward2}.

 It is handy to discuss the properties of trapped surfaces rather than event horizons, since the trapped surfaces are locally defined and can be calculated at any time during the evolution, while defining event horizons requires information about the global causal structure of the spacetime. Since a host of different initial configurations of matter might eventually evolve into the same black hole, it is the trapping horizon which contains all the interesting information concerning the process of collapse.

The Penrose inequality provides an upper bound on the area of an evolving apparent horizon $A_{H}$, given the total mass $M_{ADM}$ of the spacetime [1-7]:
\begin{eqnarray}
M_{ADM}\geq  \sqrt{\frac{A_{H}}{16 \pi}} \; . \nonumber 
\end{eqnarray} This inequality is useful since both $A_{H}$ and $M_{ADM}$ can be calculated using only the given initial information on a spacelike hypersurface. Alternatively, generating counterexamples to this potential physical ``law'' does not require extensive numerical simulation.

This inequality has been \emph{proven} in a couple of special circumstances: in the case where the spacetime is spherically symmetric \cite{malec2} \cite{iriondo}\cite{Hayward}, and in the \emph{time symmetric} (zero extrinsic curvature) case  \cite{Huisken}, \cite{mars06}, \cite{Bray}. The search for violations of the Penrose inequality must therefore focus on spacetimes that are neither. The simplest compact, non-spherical horizon we could consider would be a prolate spheroid. The idea that prolate collapse could be the key to violating the Penrose inequality or cosmic censorship is commonly attributed to Thorne \cite{Thorne} (as a violation of the Hoop Conjecture) and has been gaining popularity \cite{malechoop}.

Barrab\`{e}s, et al. \cite{barrabes} explored the subject by building models consisting of cylindrical or prolate null shells collapsing onto a Minkowski vacuum; and then determining the conditions in which the outer surface of the shell would become a marginally trapped surface. In this way, they constructed prolate, rectangular, and `puck' shaped apparent horizons. All of their models satisfied the Penrose inequality. Gibbons \cite{gibbons} later showed that their results were generally true for this type of setup (see also \cite{pelath}).   

Jaramillo, Vasset and Ansorg \cite{jaramillo}  perturbed the Kerr solution and examined the effects upon the ratio between the mass and the area of the apparent horizon.  Karkowski and Malec \cite{karkowski2} examined  prolate and oblate black holes on conformally flat hypersurfaces. The Penrose inequality was consistently satisfied by all models investigated. 

 Finally, Karkowski, Malec and \'{S}wierczy\'{n}ski \cite{karkowski} looked  at conformally and asymptotically flat \emph{prolate} or \emph{oblate} metrics on the spacelike hypersurface $\Sigma$. They then constrained their hypersurface to have zero extrinsic curvature, and examined the resulting apparent horizons. The Penrose inequality was satisfied for all of their models, a finding consistent with the proof of the Penrose inequality for time symmetric initial data \cite{Huisken}.

In this paper, we attempt to find counterexamples to the Penrose inequality by supposing first that the apparent horizon of a black hole has a prolate spheroidal shape; and then, that our initial geometry has a non-zero extrinsic curvature. Once we have constructed a solution which violates the Penrose  inequality, we show that while the dominant energy condition is satisfied about the equator, it is violated at the poles.

\section{\label{sec:level2}Background}
\subsection{Hamiltonian Formalism}
General relativity can be rewritten in the hamiltonian formalism in terms of a spacelike 3-surface $\Sigma$ with a metric  $g_{ij}$, a second fundamental form $K^{ij}$, an initial energy density $\rho$ and a momentum flux $J^{i}$. These four objects are then evolved in time, resulting in a 4-dimensional spacetime. The initial data is constrained by the Einstein Constraint Equations:
\begin{eqnarray}
R^{3}=16 \pi \rho +K^{i j}K_{ij} -(K^{i}_{i})^2   \label{eqn:rho} \\
-8\pi J^{k} = K^{i k}_{;i} -K^{i}_{i;j}g^{jk} \; . \label{eqn:J}
\end{eqnarray}

Though we are concerned with black holes and their event horizons: event horizons are inconvenient objects to work with. This is because event horizons are defined as the boundary between asymptotic future null infinity $\ell^{+}$ and the interior of the black hole \cite{wald1}. As a result of their definiton in terms of the global causal properties of a solution, locating them can require labourious numerical simulations. For convenience, we instead consider locally defined structures which can be located from the data on a 3-surface, namely:  \emph{trapped surfaces}, \emph{marginally trapped surfaces}, and \emph{apparent horizons}.

Upon our 3-surface $\Sigma$, a compact 2-surface $S_{t}$ is said to be a \emph{trapped surface} if the expansion $\theta$ of both the in-going and out-going null geodesics normal to $S_{t}$ are negative: $\theta_{\pm}<0$. That is to say, the lightcones of all points upon $S_{t}$ will end up converging. Since not all compact 2-surfaces on $\Sigma$ will be \emph{trapped}, there must be a surface $S_{o}$ upon which the expansion of the congruence of outgoing null geodesic normal to $S_{o}$ has zero outward expansion $\theta_{+}=0$. We call  surface $S_{o}$ \emph{marginally (outer) trapped}. Finally, we describe the part of the spacetime which is trapped as being the interior of the black hole, and call the \emph{outermost} marginally trapped surface the \emph{apparent horizon} -- see for example \cite{bendov}. 

Since the apparent horizon can only expand, it will eventually become the event horizon of the black hole. Alternatively, the size of the apparent horizon can act as a lower bound to the size of the eventual event horizon.

Usually, we define the expansion of the null congruence emerging from a 2-surface $S$ in terms of the tangents to the null congruence $n_{a}$. The condition for a marginally \emph{outer} trapped surface is then that $\theta_{+}=n^{a}_{;a}=0$. In terms of the spatial 3-metric $g_{ij}$ and the second fundamental form $K_{ij}$ of our initial 3-surface, this can be written \cite{husain}:
\begin{eqnarray}
\theta_{+}=(g^{ij} -n^{i}n^{j})(K_{ij} +n_{j;i})=0 \; . \label{eqn:expansion}
\end{eqnarray}

Finally we are interested in the satisfaction of the Dominant Energy Condition (DEC).Usually the DEC is defined \cite{Hawking} as the requirement that the energy density be positive and that nothing is travelling super-luminally as seen by any observer. In terms of the stress energy tensor $T_{ab}$ this can be re-written: $T_{ab}n^{a}T^{bc}n_{c}\leq0$, $T_{ab}n^{a}n^{b}\geq0$ for all timelike $n^{a}$.   In the hamiltonian formulation, the DEC can be rewritten in terms of energy density $\rho$ and the energy flux vector $J^{i}$ as seen by an observer on the initial surface:
\begin{eqnarray}
\rho \pm |J| \geq 0 \; . \label{eqn:DEC}
\end{eqnarray}

\subsection{Asymptotic Flatness and the ADM Mass}

 We say that the spacelike 3-surface $\Sigma$ is  \emph{asymptotically flat} if: \emph{i.} $\Sigma$ is the disjoint union of a compact set,  and a set diffeomorphic to $R^{3}\setminus B$ where $B$ is a closed ball \cite{mars06}; and  \emph{ii.} the spatial metric $g_{i j}$ on $\Sigma$ and the second fundamental form $K_{i j}$ fall off in the radial coordinate $r$ as:

\begin{eqnarray}
g_{i j} = \delta_{i j} + O(\frac{1}{r}) \nonumber \\
K_{i j} = O(\frac{1}{r^{2}}) \nonumber \; \; .
\end{eqnarray} 

The first constraint (\emph{i.}) is a statement concerning the global structure of the spacetime: there exists a ``spacelike infinity'' (the compact set) and a spacelike manifold which can contain a black hole (the set diffeomorphic to $R^{3}\setminus B$). The second constraint (\emph{ii.}) ensures that the Ricci Scalar decays as $R=O(\frac{1}{r^{4}})$, which is sufficient to ensure that the ADM mass is a geometric quantity.

The Arnowitt-Deser-Misner (ADM) mass is a equivalent to the total rest-mass of the energy in the spacetime \cite{geroch}. For instance,  it matches up with the black hole mass in the Schwarzchild spacetime.

 It is evaluated by taking the limit of the following integral over the area of a sphere $S_{\sigma}$ of constant radius $\sigma$ with area element $d\mu$ and normal vector $v_{j}$  defined on a spacelike hypersurface $\Sigma$ \cite{Bray}:

\begin{eqnarray}
M_{ADM}=\frac{1}{16 \pi} \lim_{\sigma \to \infty} \int_{S_{\sigma}} \Sigma_{i,j} (g_{ij,i} v_{j} -g_{ii,j}v_{j}) d\mu \nonumber
\end{eqnarray}

If the spacetime is \emph{asymptotically flat}, the ADM mass will be invariant in time and under different hypersurface slicings.

\section{The Penrose Inequality}

\subsection{Cosmic Censorship}

The Cosmic Censorship Conjecture (CCC) \cite{penrose2} says, simply, that the causal past of future null infinity $J^{-}[\ell^{+}]$ must be geodesically complete. Alternatively put, any singularities in the spacetime must be surrounded by event horizons. 

Numerical relativity has sought to test this conjecture by constructing models which collapse into singularities in some ``realistic'' way that might then serve as counterexamples \cite{wald1}, [20-23]. While this procedure for exploring (or disproving) the CCC is interesting, it is intricate and difficult. One could alternatively test the CCC by trying  to find physically realizable counter-examples to the Penrose inequality \cite{penrose1}. 
 
\subsection{The Penrose Inequality}

 Consider an asymptotically flat spacelike 3-surface $\Sigma$ upon  which some matter is in the process of collapsing into a black hole, and suppose additionally that the matter is physically reasonable (the DEC is satisfied throughout). 

For Schwarzchild data, the area of the event horizon $A_{EH}$ can be related to the Schwarzchild mass $M_{BH}$: 
\begin{eqnarray}
A_{EH} =4 \pi (2M_{BH})^{2} \nonumber \; .
\end{eqnarray}
If  $\Sigma$ is an asymptotically flat slice with matter, then $M_{BH}< M_{ADM}$ ($M_{BH}=M_{ADM}$ exclusively in the static Schwarzchild case). (Note: this assumption uses the \emph{Positive Mass Theorem}, which relies on the CCC.)

An evolving black hole can be described in terms of an expanding apparent horizon with area $A_{H}$. As the black hole accretes the matter around it, its apparent horizon will grow until there is no longer any mass for the black hole to accrete and it becomes an event horizon $A_{EH}$.  Therefore: $A_{H}\leq A_{EH}$. 

Combining these, we end up with the inequality  \cite{penrose1}:
\begin{eqnarray}
M_{ADM}^2 \geq   \frac{A_{H}}{16 \pi} \; . \nonumber
\end{eqnarray}

Let us define the \emph{surface of smallest area} $S_{o}$ which \emph{contains} the apparent horizon $A_{H}$.  Unless $\Sigma$ has been endowed with a strange geometry, $S_{o}$ and $A_{H}$ will be the same surface. 

If $A_{S_{o}}$ is the area of $S_{o}$, then our inequality takes on its most general form: the Penrose inequality
\begin{eqnarray}  \label{eqn:penroseinequality}
M_{ADM}\geq \sqrt{\frac{A_{S_{o}}}{16 \pi}} \; .
\end{eqnarray}
The Penrose inequality additionally specifies that the equality only holds in the case of the Schwarzchild solution.

Recall that this inequality depended upon three postulates: asymptotic flatness, the satisfaction of the DEC, and the CCC. Consequently, if one could find an asymptotically flat solution which satisfies the DEC but violates the inequality (\ref{eqn:penroseinequality}): one would have found a counterexample the CCC \cite{wald1}.

The Penrose inequality has already been proven for certain specific circumstances without requiring the CCC. The first proof, by Hayward \cite{Hayward}, \cite{mars06}, \cite{Bray} assumes that the shape of the horizon, the first and the second fundamental form are all spherically symmetric. The second proof, by Huisken and Ilmanen \cite{Huisken}, \cite{mars06}, \cite{Bray} assumes that the initial hypersurface has zero extrinsic curvature (this is frequently called the \emph{time symmetric} case). Consequently, if a physically realizable counterexample to the Penrose inequality even exists, it must not be spherically symmetric and its second fundamental form  must be non-zero.

\section{\label{mywork}Constructing a Prolate Apparent Horizon Violating the Penrose Inequality}

Our work follows a reverse approach to the problem than those reviewed in section \ref{sec:level1}: we  first construct initial data with prolate apparent horizons that violate the Penrose inequality, and then we determine whether or not our solution satisfies the DEC. 

We begin our construction by defining an orientable spacelike hypersurface that will have an asymptotically flat prolate geometry, and a second fundamental form $K_{ij}$, which dies off appropriately.

\begin{enumerate}

\item We fix a prolate metric with mass $M$ and a surface $S$ such that its area $A_{S}$ satisfies $M = \sqrt{\frac{A_{s}}{16\pi}}$. This will violate the Penrose inequality, since the geometry is not spherically symmetric.

\item We require that the surface $S$ be an apparent horizon by forcing $\theta_{+}=0$, $\theta_{-}<0$ for the null geodesics emerging from it. We do so by fixing the freedom available in $K_{ij}$.

\item We check to see whether our solution satisfies the DEC upon the apparent horizon. This is a necessary but not sufficient condition for physicality. 

\end{enumerate}

Following Karkowski \cite{karkowski}, we define our asymptotically (and conformally) flat metric, in prolate spheroidal coordinates. \begin{eqnarray}
ds^2 =\left(1+\frac{M}{2\sigma} \right)^4 \left(  \frac{\sigma^2-\tau^2}{\sigma^2-1} d\sigma^2 \right. \; \; \; \; \; \; \; \; \; \; \; \; \; \; \label{metric} \\
\left. + \frac{\sigma^2 -\tau^2}{1-\tau^2} d\tau^2 +(\sigma^2 -1)(1-\tau^2)d\phi^2  \right) \;.
\nonumber
\end{eqnarray}

\subsection{\label{area}Defining the  Apparent Horizon on $\Sigma$}

We define the apparent horizon on $\Sigma$ to be a 2-surface of constant coordinate radius: $ \sigma=\hat{\sigma} $. Due to the way prolate spheroidal coordinates are defined, as $\hat{\sigma}\rightarrow 1$, the degree to which our horizon is distended will increase. 

Given the metric, the area of such a surface is:
\begin{eqnarray}
A=\int_{-1}^{1} \int_{0}^{2 \pi} \left(   1+\frac{M}{2 \hat{\sigma}} \right) ^{4} \sqrt{ \hat{\sigma}^2 -\tau^2} \sqrt{\hat{\sigma}^2 -1} d\phi d\tau 
\\
= 2 \pi  \left(   1+\frac{M}{2 \hat{\sigma}} \right) ^{4}  \sqrt{\hat{\sigma}^2 -1}
 \left( \sqrt{\hat{\sigma}^2 -1} +\hat{\sigma}^2 \sin^{-1}{ \frac{1}{\hat{\sigma}} } \right) \; .
\label{eqn:area}
\end{eqnarray}

Given an apparent horizon of radius $\hat{\sigma}$, we can fix the ADM mass in order to violate the Penrose inequality using:

\begin{eqnarray}
8 M^{2} =  \left(   1+\frac{M}{2 \hat{\sigma}} \right)^{4} \sqrt{\hat{\sigma}^2 -1}
 \left( \sqrt{\hat{\sigma}^2 -1} +\hat{\sigma}^2 \sin^{-1}{\frac{1}{\hat{\sigma}}} \right) \; .
\label{eqn:horizonradius}
\end{eqnarray}

For a surface of constant $\sigma=\hat{\sigma}$ to be an apparent horizon, its second fundamental form  must be constrained so that none of the families of  null geodesics emerging from it have positive expansion.

 The unit normal of our  2-surface of constant radius will be:
\begin{equation}
n^{i}=\left[  \frac{4 \hat{\sigma}^2}{\sqrt{\frac{(2 \hat{\sigma} +M)^{4} (\hat{\sigma}^2-\tau^2)}{\hat{\sigma}^2 -1}}}    ,0,0     \right] \; \; .
\label{eqn:normal}
\end{equation}

Additionally, we note that the Penrose inequality requires that our hypersurface be asymptotically flat. Thus we will assume that the extrinsic curvature is defined using parameters $a,b,c$ and  has the form:

\begin{eqnarray}
K_{ij} =\left[ 
 \begin{array}{c} \frac{a}{\sigma^2} \\ \frac{b}{\sigma^2 \sqrt{1-\tau^2}} \\ 0 \end{array}    ,   
 \begin{array}{c} \frac{b}{\sigma^2 \sqrt{1-\tau^2 }} \\ \frac{c}{\sigma^2 (\tau^2-1))} \\ 0 \end{array}   ,\begin{array}{c} 0 \\ 0 \\ 0 \end{array}                \right]
\label{eqn:formextrinsic}
\end{eqnarray}

The parameters $a$ and $c$ are related to the invariants:
\begin{eqnarray}
a= -\frac{1}{16} K_{a b}n^{a}n^{b}\frac{(2 \sigma +M)^{4} (\tau^2 -\sigma^2)}{\sigma^2 (\sigma^2-1)} \nonumber \\
c= \frac{1}{16} (K_{a b}g^{ab}-K_{a b}n^{a}n^{b}) \frac{(2 \sigma +M)^{4} (\tau^2 -\sigma^2)}{\sigma^2 } \nonumber
\end{eqnarray}
Thus, if $a$ and $c$ remain finite, so will the invariants $K^{a b}n_{a}n_{b}$ and $K_{ab}g^{ab}$ ($b$ can be related in a similar way to $K^{a b}K_{a b}$).  

To make our prolate surface  surface marginally trapped ($\theta_{+}=0$) (\ref{eqn:expansion}), we constrain one of the terms in our second fundamental form:

\begin{eqnarray}
c &=& \left( \frac{2\hat{\sigma}+M}{4\hat{\sigma} \sqrt{ (\hat{\sigma}^2 -\tau^2)(\hat{\sigma}^2 -1)} } \right) \times \; \; \;  \\
 ( 2 &M&\hat{\sigma}^4 - 4 \hat{\sigma}^5  + 4M\tau^2 +(2 \hat{\sigma}^3 -3M \hat{\sigma}^2) (1+\tau^2)  )  \nonumber \; .
\label{eqn:c} 
\end{eqnarray}

The other set of null geodesics emerging from the marginally trapped surface $\sigma=\hat{\sigma}$ must converge ($\theta_{-}<0$) for our data to represent a black hole. We therefore require that:

\begin{eqnarray}
\theta_{-}=(g^{ij} -n^{i}n^{j})(K_{ij} -n_{j;i})<0 \; . \label{eqn:expansionelle}
\end{eqnarray}

\begin{figure}
\includegraphics[width=8cm]{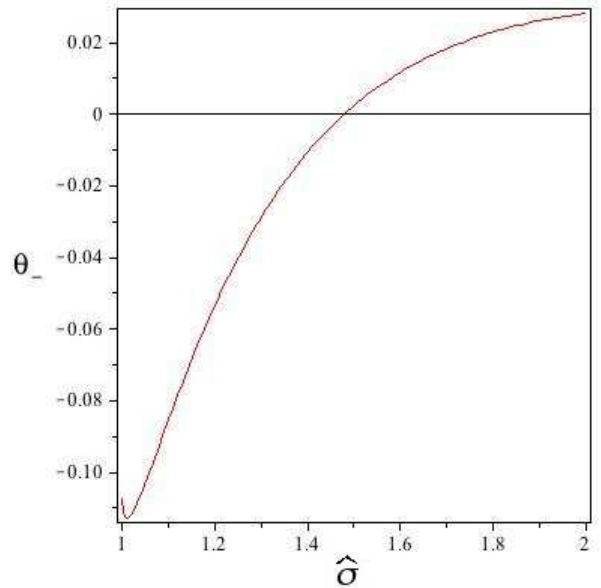}
\caption{\label{fig:blackholewhitehole}   $\theta_{-}$ is plotted as a function of the radius of the marginally trapped surface $\hat{\sigma}$. The surface will be outer-trapped  and our data will represent a black hole when $\theta_{-}<0$. 
}
\end{figure}

We plot $\theta_{-}$, as a function of the radius of the marginally trapped surface $\sigma=\hat{\sigma}$ in figure (\ref{fig:blackholewhitehole}): when it is negative, the surface will be outer-trapped, and our data will represent a black hole. From the graph, we conclude that we should only consider trapped surfaces with $1<\hat{\sigma}<1.45$. 

\subsection{\label{DEC}The Dominant Energy Condition}

Since $R$ and $K^{ij}$ have now been defined, $\rho$ and $J$ will be defined through equation (\ref{eqn:rho}) and we can now determine whether the DEC is satisfied.  We calculate $|J|=\sqrt{J^{i} J_{i}}$, and plot $\rho-|J|$ and $\rho$ across the surface $\sigma=\hat{\sigma}$.

In addition to the radius $\hat{\sigma}$, and the angle $\tau$; $\rho$ and  $\rho-|J|$ will depend on the undefined parameters in $K_{ij}$: $a$, $b$ which we will assume are constant on surface $\sigma=\hat{\sigma}$. Let us consider what values of $a$ and $b$ are most likely to satisfy the DEC.

\begin{figure}
\includegraphics[width=8cm]{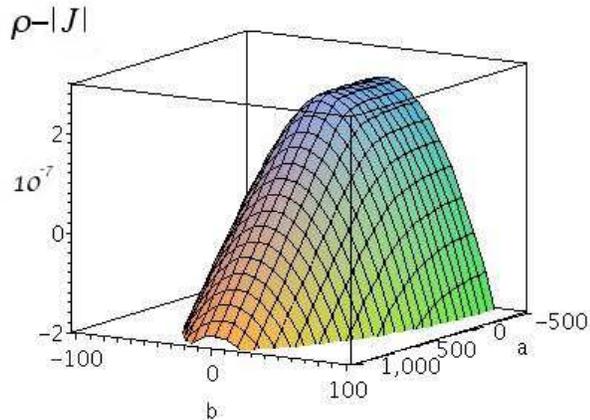}
\caption{\label{fig:ab} In this diagram we look at $\rho-|J|$ on the equator of an radius $\hat{\sigma}=1.4$ black hole, as a function of the parameters $a$ and $b$ from $K_{ij}$. From this we set $a \ll 0$ and $b$ small. }
\end{figure}

The graph (\ref{fig:ab}) shows how $\rho-|J|$ depends on $a$ and $b$ on the equator of a radius $\hat{\sigma}=1.4$ black hole slice. We see that $\rho-|J|$ increases and becomes positive as $a$ is increasingly negative; while $b$ should be set to nearly zero. We set $a=-100$, and $b=2$.

Let us now look at whether the DEC is satisfied upon the apparent horizon, and whether changing the size of the horizon  has any effect on its satisfaction.

\begin{figure}
\includegraphics[width=8cm]{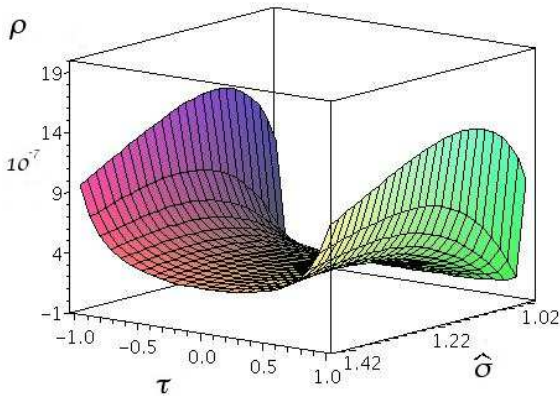}
\caption{\label{fig:faroutrho} $\rho$ for all values of angle $\tau$ on the surface of the horizon where $a=-100$, $b=2$ for a variety of radii $\hat{\sigma}$. This plot shows that the energy density on the horizon is generally positive for the solutions we are considering.}
\end{figure}

 In plot (\ref{fig:faroutrho}) we show that the energy density $\rho$ is positive everywhere upon the horizon ($-1 \leq \tau \leq 1$), for a variety of horizon radii $\hat{\sigma}$.

\begin{figure}
\includegraphics[width=8cm]{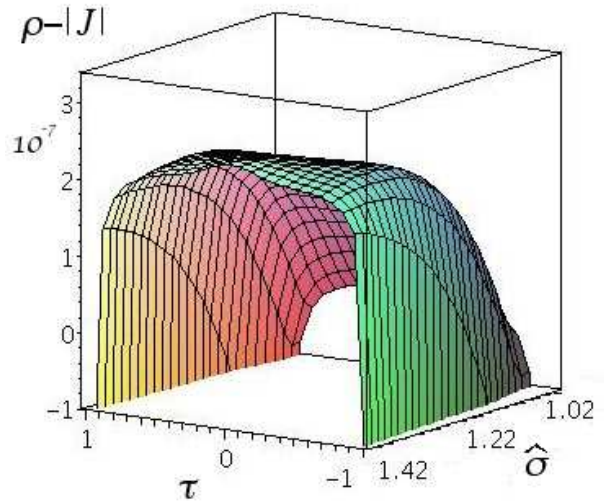}
\caption{\label{fig:farout} $\rho-|J|$ for all values of angle $\tau$ on the surface of the horizon where $a=-100$, $b=2$ for a variety of possible radii $\hat{\sigma}$. This plot shows that the DEC violations will remain near the poles. }
\end{figure}

 The plot (\ref{fig:farout}) of  $\rho-|J|$ as a function of $\tau$ and $\hat{\sigma}$  demonstrates that the DEC can be satisfied around the equator of the horizon, but that it will always be violated at the poles ( $\tau \rightarrow \pm 1$) where $\rho-|J|$ becomes negative and diverges. This violation occurs regardless of the radii $\hat{\sigma}$ of the apparent horizon.

\begin{figure}
\includegraphics[width=8cm]{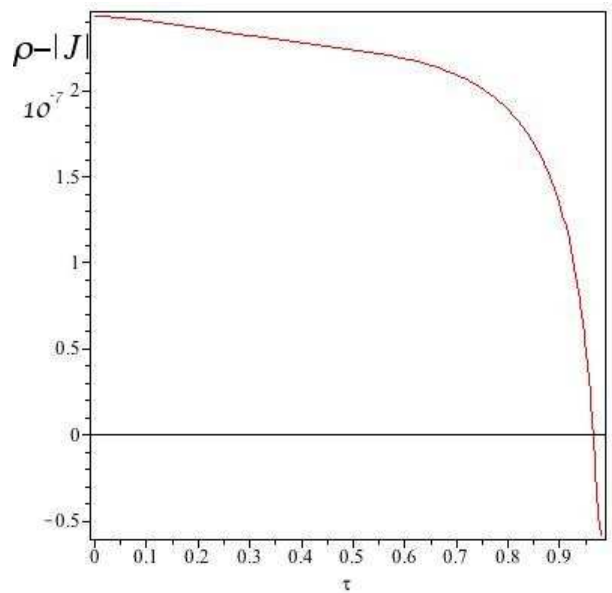}
\caption{\label{fig:crossection}  $\rho-|J|$ for all values of angle $\tau$ on the surface of the horizon, for horizon at $\hat{\sigma}=1.4$, $a=-100$, $b=2$. The negative divergence at the poles indicate that the DEC has been violated.}
\end{figure}

Consider as a specific model: A low mass, highly prolate apparent horizon ($\hat{\sigma}=1.4$), which satisfies the DEC in the tropics, and violates it near the poles (figure (\ref{fig:crossection})).

\section{\label{final}Discussion}
Our results rule out this class of non-symmetric data as potential counterexamples of the Penrose inequality, due to the violation of the DEC. The violations will occur at the north and south pole of the apparent horizon; and will occur regardless of how severe or how slight the prolate warping of the spheroid is (since it occurs for the range of $\hat{\sigma}$).

It should be noted that while the specific violations of the DEC around the poles are due to the way we specified the second fundamental form $K_{a b}$; similar violations of the DEC occurred when we specify $K_{a b}$ in different ways. Note that since our construction is coordinate-dependant, we needed to be wary of the coordinate singularities at $\tau=\pm 1$. We tried a variety of approaches to resolving this issue. An alternate (but more cumbersome) approach is to define the individual components of $K_{a b}$ directly in terms of geometric invariants, which we would like to remain finite. In every other method we explored in defining $K_{a b}$, however, the DEC was violated somewhere. We chose (\ref{eqn:formextrinsic}) because it is algebraically the simplest and required dramatically fewer computational resources; and also because there will remain large regions of the horizon where the DEC is satisfied.

\section{Acknowledgments}
This work was supported in part by the Natural Sciences and Engineering Research Council of Canada.  The author is extremely grateful to Viqar Husain for his guidance throughout this work.

\bibliography{prolatepaper4}

\end{document}